\documentclass[twocolumn,showpacs,preprintnumbers,amsmath,amssymb]{revtex4}
\usepackage{graphicx}% Include figure files
\input epsf
\usepackage{dcolumn}% Align table columns on decimal point
\usepackage{bm}% bold math

\begin{document}

%\preprint{Theoretical and Mathematical Physics, 139 (3): 787-800 (2004)}

\title{Quantum Creation of a Universe with
 varying speed of light: $\Lambda$-problem and Instantons}% Force line breaks with \\
\author{A.V. Yurov}
\email{artyom_yurov@mail.ru}
%\altaffiliation[Also at ]
{%
\affiliation{%
I. Kant Russian State University, Theoretical Physics Department,
 Al.Nevsky St. 14, Kaliningrad 236041, Russia
\\
 }
\author{V.A. Yurov}%
 \email{valerian@math.missouri.edu}
 \affiliation{%
Department of Mathematics, University of Missouri, Columbia, MO
65211, U.S.A.
\\
}%

%\affiliation{

\date{\today}% It is always \today, today,
\begin{abstract}
One of the most interesting development trends of a modern
cosmology is the analysis of models of a modified gravitation.
Without exaggeration it is possible to say that Sergei Odintsov is
one of the leaders of this direction of researches (see
\cite{Odin}). This article is dedicated to cosmologies with
variable speed of light (VSL) - models, which one can consider as
a particular case of models of a modified gravitation.

In quantum cosmology the closed universe can spontaneously
nucleate out of the state with no classical space and time. The
semiclassical tunneling nucleation probability can be estimated as
$\emph{P}\sim\exp(-\alpha^2/\Lambda)$ where $\alpha$=const and
$\Lambda$ is the cosmological constant.

In classical cosmology with varying speed of light $c(t)$  it is
possible to solve the horizon problem, the flatness problem and
the $\Lambda$-problem if $c=sa^n$ with $s$=const and  $n<-2$. We
show that in VSL quantum cosmology with $n<-2$ the semiclassical
tunneling nucleation probability is
$\emph{P}\sim\exp(-\beta^2\Lambda^k)$ with $\beta$=const and
$k>0$. Thus, the semiclassical tunneling nucleation probability in
VSL quantum cosmology is very different from that in quantum
cosmology with $c$=const. In particular, it can be strongly
suppressed for large values of $\Lambda$. In addition, we propose
two instantons that describe the nucleation of closed universes in
VSL models. These solutions are akin to the Hawking-Turok
instanton in sense of $O(4)$ invariance but, unlike to it, are
both non-singular. Moreover, using those solutions we can obtain
the probability of nucleation which is suppressed for large value
of $\Lambda$ too. We also discuss some unusual properties of
models with $n>0$.

\end{abstract}

\pacs{98.80.Cq;98.80.-k}% PACS, the Physics and Astronomy
\maketitle

\section{\label{sec:level1}Introduction}

One of the major requests concerning the quantum cosmology is a
reasonable specification of initial conditions in early universe,
that is in close vicinity of the Big Bang. There are known the
three common ways to describe quantum cosmology: the
Hartle-Hawking wave function \cite{4}, the Linde wave function
\cite{5}, and the tunneling wave function \cite{6}. In the last
case the universe can tunnel through the potential barrier to the
regime of unbounded expansion. Following Vilenkin \cite{7} lets
consider the closed ($k=+1$) universe filled with radiation
($w=1/3$) and $\Lambda$-term ($w=-1$). One of the Einstein's
equations can be written as a law of a conservation of the
(mechanical) energy: $P^2+U(a)=E$, where $P=-a\dot a$, $a(t)$ is
the scale factor, the ''energy'' $E=$ const and the potential
$$
U(a)=c^2a^2\left(1-\frac{\Lambda a^2}{3}\right),
$$
where $c$ is the speed of light; see Fig.1
%%%%%%%%%%%%%%%%%%%%%%%%%%%%%%%%%%%%%%%%%%%%%%%
\begin{figure}
\centering\leavevmode\epsfysize=5.5cm \epsfbox{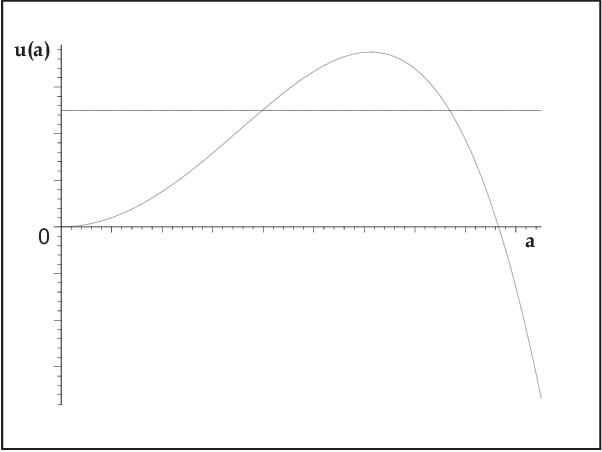}
\newline
\caption{\label{fig:epsart} Vilenkin potential with $c=$const.}
\end{figure}
%%%%%%%%%%%%%%%%%%%%%%%%%%%%%%%%%%%%%%%%%%%%%%%%%%%
The maximum of the potential $U(a)$ is located at
$a_e=\sqrt{3/2\Lambda}$ where $U(a_e)=3c^2/(4\Lambda)$. The
tunneling probability in WKB approximation can be estimated as
\begin{equation}
\emph{P}\sim\exp\left(-\frac{2c^2}{8\pi G\hbar}\int_{a'_i}^{a_i}
da\sqrt{U(a)-E}\right), \label{1}
\end{equation}
where $a'_i<a_i$ are two turning points. The universe can start
from $a=0$ singularity, expand to a maximum radius $a'_i$ and then
tunnel through the potential barrier to the regime of unbounded
expansion with the semiclassical tunneling probability (\ref{1}).
Choosing $E=0$ one gets $a'_i=0$ and $a_i=\sqrt{3/\Lambda}$. The
integral in (\ref{1}) can be calculated. The result can be written
as
\begin{equation}
 \emph{P}\sim
\exp\left(-\frac{2c^3}{8\pi G\hbar\Lambda}\right). \label{2}
\end{equation}
For the probability to be of reasonable value, for example
$\emph{P}=1/{\rm e}\sim 0.368$, one has to put $\Lambda\sim
0.3\times 10^{65}\,\, {\rm cm}^{-2}$ (see (\ref{2})). In other
words, the $\Lambda$-term must be large. However, despite this
problem, we does acquire one prise: Once nucleated, the universe
immediately begins a de Sitter inflationary expansion. Therefore
the tunneling wave function results in inflation. And the
$\Lambda$-term problem, which arises in this approach is usually
being gotten rid of via the anthropic principle (another way to solve $\Lambda$-problem is  brane-worlds models
\cite{LeraTema}).
%%%%%%%%%%%%%%%%%%%%%%%%%%%%%%%%%%%%%%%%%%%%%%%%%%%%%

Now let's return to the Fig. 1.  We have two Lorentzian regions
($0<a<a'_i$, $a>a_i$) and one Euclidean region ($a'_i<a<a_i$). The
second turning point $a=a_i$ corresponds to the beginning of our
universe. If $\Lambda=0$ then $U(a)$ has the form of {\bf
parabola} and we get only one Lorentzian region. In this case, the
universe can start at $a=0$, expand to a maximum radius  and then
recollapse. If $E\to 0$ then the single Lorentzian region is
contract to the point. This, of course, comes to an agreement with
the tunnelling nucleation probability: $\emph{P}\to 0$ as
$\Lambda\to 0$. In this article, however, we'll show that in
quantum cosmological VSL models  the situation can be opposite,
viz: the probability to find the finite universe short after it's
tunneling through the potential barrier is
$\emph{P}\sim\exp(-\beta(n)\Lambda^{\alpha(n)})$ with
$\alpha(n)>0$ and $\beta(n)>0$ when $n<-2$ or for $-1<n<-2/3$.
After the tunneling one get the finite universe with "initial"
value of scale factor $a_i\sim\Lambda^{-1/2}$, so the probability
to find the universe with large value of $\Lambda$ and small value
of $a_i$ is strongly suppressed. The reason of this lies in the
behavior of potential $U(a)$, which, for the case $\Lambda\to 0$,
is being transformed into the {\bf hyperbola}, located under the
abscissa axis. As a result, such a universe can start at $a\sim 0$
the regime of unbounded expansion. Therefore, we get the single
Lorentzian region which is not contract to the point at $E\to 0$.

This new property of VSL quantum cosmology will be discussed in
the Sec. II for the case $w=1/3$. But there arouse the two new
questions which have to be answered. First at all, the effective
potential in VSL models can be unbounded from below at $a\to 0$.
What possible meaning of such potentials can be? The second
question is the geometric interpretation of the quantum creation
of a Universe with varying speed of light. We know that universe
can be spontaneously created from nothing (in model where $c={\rm
const}$) and this process can be described with the aid of the
instantons solutions possessing  $O(5)$ (if $V(\phi)$ has a
stationary point at some nonzero value $\phi=\phi_0={\rm const}$)
or $O(4)$ (as Hawking-Turok instanton \cite{HT}) invariance. So,
what can be said about instantons in the VSL models?

The whole plan of the paper looks as follows: in the next Section
we'll consider the simplest VSL model: model of
Albrecht-Magueijo-Barrow. Then we show that in framework of
tunneling approach to quantum cosmology with VSL the semiclassical
tunneling nucleation probability can be estimated as
$\emph{P}\sim\exp(-\beta^2\Lambda^k)$ with $\beta$=const and
$k>0$. All corresponding calculations will be done for the case of
the universe filled with radiation ($w=1/3$) and vacuum energy.
The case of any $w$ will be considered in the Sec. III, where we
will discuss the problem of potentials unbounded from below at
$a\to 0$. We'll use the following naive procedure: By analogy with
Schr\"odinger equation we apply the Heisenberg uncertainty
relation to the Wheeler-DeWitt equation in order to find the
potentials which admit the ground state, even being unbounded from
below at $a\to 0$.

In the Section IV we'll propose the {\bf non-singular} instanton
solutions possessing only $O(4)$ invariance (so the Euclidean
region is a deformed four sphere). These solutions can in fact
lead to inflation after the analytic continuation into the
Lorentzian region. We will discuss these results in Sec. V. Some
unusual properties of VSL models with $n>0$ (including ''big rip''
and ''big trip'') are discussed in Appendix.
%%%%%%%%%%%%%%%%%%%%%%%%%%%%%%%%%%%%%%%%%%%%%%%%%%%%%%%%%
\section {Albrecht-Magueijo-Barrow VSL model}
Lets start with the Friedmann and Raychaudhuri system of equations
with $k=+1$ (we assume the $G$=const):
\begin{equation}
\begin{array}{cc}
\displaystyle{\frac{\ddot a}{a}=-\frac{4\pi
G}{3}\left(\rho+\frac{3 p}{c^2}\right)+\frac{\Lambda c^2}{3},}
\\
\\
\displaystyle{\left(\frac{\dot a}{a}\right)^2=\frac{8\pi
G\rho}{3}-k\left(\frac{c}{a}\right)^2+\frac{\Lambda c^2}{3},}
\\
\\
\displaystyle{c=c_0\left(\frac{a}{a_0}\right)^n=sa^n,\qquad
p=wc^2\rho,}
\end{array}
\label{frid}
\end{equation}
where $a=a(t)$ is the expansion scale factor of the Friedmann
metric, $p$ is the fluid pressure, $\rho$ is the fluid density,
$k$ is the curvature parameter (we put $k=+1$), $\Lambda$ is the
cosmological constant, $c_0$ is some fixed value  of speed of
light which corresponds some fixed value of scale factor  $a_0$.

{\bf Remark 1}. In \cite{Ellis} George F.R. Ellis and
Jean-Philippe Uzan has pointed out at the fact, that  the system
of Friedmann-Raychaudhuri equations (system (\ref{frid})), as
denoted in our article) is not consistent from the point of view
of a field theory. This is because the system (\ref{frid})
couldn't be derived via the standard variation procedure, despite
the contrary claims, given in [7]-[10] (see the references in
\cite{Ellis}). Variation of action (36) (see \cite{Ellis}) leads
instead to the system of equations (39)-(42) (see \cite{Ellis}),
and not to the Friedmann-Raychaudhuri. Therefore, since our
article is initially based on (\ref{frid}) it seems that the
present article is just a formal exercice based on unreliable
grounds (Eqs. (\ref{frid})) so that its conclusions cannot tell us
much about the effect of varying constants in Quantum cosmology.

We don't think so. The fact, that the equations (39)-(42) (of
\cite{Ellis}) are distinct from set (\ref{frid}) doesn't
immediately mean that the system (39)-(42) (of  \cite{Ellis}) is
correct while (\ref{frid}) is not. Moreover, even assuming that
George F.R. Ellis and Jean-Philippe's claim is indeed true and it
is impossible to get (\ref{frid}) from (36) (see  \cite{Ellis}),
this would not have any consequences for our further results,
barring the instanton chapter. In our article (except the
mentioned chapter) we use neither the action (36) (of
\cite{Ellis}) nor the fact that (\ref{frid}) can be derived from
(36). Contrary to that, we just use (\ref{frid}) as a basic
phenomenological model. This, of course, can be regarded as a flaw
of a model, but same can be said about the system (39)-(42) (of
\cite{Ellis}), which is also but a phenomenological model, based
on assumption, that the VSL models are in essence the particular
example of the scalar-tensor theories (see for example
\cite{Odin1}).

Hence, in order to be able to make the preference of (39)-(42) (of
\cite{Ellis}) over equation (\ref{frid}) (or vice verse) we have
to gain the better comprehension of the physical principles, lying
behind the variability of the speed of light (in assumption, that
it really changes at all, which is still far from obvious!..).
That, in turn, would be possible only after we will understand the
origins of "c". It has been shown in  reviewing part of
\cite{Ellis}, that it is possible to define not just one, but a
couple of "speeds of light" ($c_{ST}$, $c_{EM}$, $c_E$). Beyond
any doubts, there should exist a fundamental physical reason for
those (generally different) values being perfectly equal. Such
reason can only be discovered in the future fundamental physical
theory (string theory?). Only such theory would verify which
equations are true: (39)-(42) (in  \cite{Ellis}), (\ref{frid}) or
maybe some other ones.

Returning to system (39)-(42) (in  \cite{Ellis}), which may be
considered as a more reliable one than Friedmann and Raychaudhuri
system (\ref{frid}), we shall in turn remind about some of the
assumptions, that has been used by George F.R. Ellis and
Jean-Philippe's in order to get to (39)-(42). For example, authors
consider the value of $c_E$ (which is used in Einstein's equation
in a form  $8\pi G/c^4$) as a variable, while assuming that the
value of $c_{ST}$, that appears in the integral
$$
ds^2=-c_{ST}^2dt^2+(dx^1)^2+(dx^2)^2+(dx^2)^2,
$$
satisfies $c_{ST}=1$ (see page. 8, in the last line above the
formula (39) in \cite{Ellis}). But if we'll assume that all
"speeds of light" are in fact the representations of just one
value (an assumption, that we strongly believe to be true), then
the value of $c=\psi^{1/4}$ will require to be multiplied by the
lagrangian ($d^4x=cdtd^3x$), including the various fields of
matter and shall also be taken into account in the multiplier
$\sqrt{g}$. With all this in mind, it becomes really doubtful that
the VSL model (36) (\cite{Ellis}) will remain being equal to some
scalar-field  theory. On the other hand, if we'll insist on
difference between the values of $c_{ST}$ and $c_E$, while
requiring the variability of both of them, it doesn't seem
impossible to choose the $c_{ST}$ in a way, that leads to results,
similar to the ones, received in [7]-[10] (see the reference list
of  \cite{Ellis}), which is sufficiently what we need for the IV
chapter of our work, devoted to instantons (and is necessary for
this chapter only!). Of course, this might look quite artificial,
but it just serves to show that both (39)-(42) of  \cite{Ellis}
and (\ref{frid}) shall be regarded as phenomenological models.

Therefore, it appears, that studying of equations (\ref{frid}) are
no more (or less) well-grounded, then studying of equations
(39)-(42) of \cite{Ellis}. As follows from our work, system
(\ref{frid}) can be quite interesting in the quantum cosmology,
since it allows for an unexpected solution of a cosmological
constant problem. This, with regard to the overall difficulty of
the problem, can be considered as an additional indication that
the model (\ref{frid}) does indeed deserve the further
examinations. We aren't making any further claims; it is of course
quite possible that the system (39)-(42) of \cite{Ellis}, if being
applied to the quantum cosmology is also capable of producing of
the interesting results, but that is a topic of a different
research.

In conclusion, let us note that the equations (39)-(42) of
\cite{Ellis} are in fact completely incompatible with our model.
Our article is based on assumption that there exist a one-to-one
correlation between $c$ and the scalar factor $a$: $c=sa^n$. In
other words, the speed light (or $\psi=c^4$ from  \cite{Ellis}) is
a function of a Ricci scalar $R$, and, hence, is NOT an
independent variable. But then we come to contradiction with the
basic assumption of  \cite{Ellis}, which one basically allowed to
get the system (39)-(42). Therefore, it would even formally be a
mistake to use the (39)-(42) of  \cite{Ellis} in a way, similar to
the one, that we adopted for (\ref{frid}) - at least without
special pre-given restrictions.

{\bf Remark 2.} One can consider more general model. Let
$$
c=sa^n+\frac{m}{a^m},
$$
with $n>0$, $m>0$, $r>0$, $s>0$. In this case the flatness and
$\Lambda$-term problems can be solved if
$$
n>m,\qquad n>1,\qquad m>\frac{3(w+1)}{2},
$$
for the weak energy condition is valid: $w>-1$. If $a\to\infty$
then the Hubble root
\begin{equation}
H^2\to\frac{\Lambda s^2(1+w)a^{2n}}{3(1+w)+2n}, \label{ola-la}
\end{equation}
so we have the big rip singularity without any phantom fields (see
Appendix A). All this can be obtained
 from the expression  $\rho$ which has the form:
\begin{widetext}
$$
\begin{array}{cc}
\displaystyle{ \rho=\frac{M}{a^{3(w+1)}}+\frac{3ks^2n
a^{2(n-1)}}{4\pi G(1+3w+2n)}+\frac{3ksr(n-m)a^{n-m-2}}{4\pi
G(1+3w+n-m)}-\frac{3kr^2m}{4\pi
G(1+3w-2m)a^{2(1+m)}}-\frac{\Lambda s^2na^{2n}}{4\pi
G(3(1+w)+2n)}+}\\
\displaystyle{ +\frac{\Lambda r^2m}{4\pi
G(3(1+w)-2m)a^{2m}}-\frac{\Lambda sr(n-m)a^{n-m}}{4\pi
G(3(1+w)+n-m)}}.
\end{array}
$$
\end{widetext}

After that extensive  digression   let's return to the equations
(\ref{frid}). Using these ones
\begin{equation}
\dot\rho=-\frac{3\dot
a}{a}\left(\rho+\frac{p}{c^2}\right)+\frac{{\dot
c}c(3-a^2\Lambda)}{4\pi G a^2}. \label{rt}
\end{equation}
Choosing $w=1/3$ one can solve (\ref{rt}) to receive
\begin{equation}
\rho=\frac{M}{a^4}+\frac{3s^2na^{2(n-1)}}{8\pi G(
n+1)}-\frac{s^2n\Lambda a^{2n}}{8\pi G(n+2)}, \label{rho}
\end{equation}
where $M>0$ is a constant characterizing the amount of radiation.
It is clear from the (\ref{rho}) that the flatness problem can be
solved in a radiation-dominated early universe by an interval of
VSL evolution if $n < -1$, whereas the problem of $\Lambda$-term
can be solved only if $n<-2$.  The evolution equation for the
scale factor $a$ (the second equation in system (\ref{frid})) can
be written as
\begin{equation}
p^2+U(a)=E, \label{equation}
\end{equation}
where $p=-a\dot a$ is the momentum conjugate to $a$, $E=8\pi G
M/3$ and
\begin{equation}
U(a)=\frac{s^2a^{2n+2}}{n+1}-\frac{2s^2\Lambda a^{2n+4}}{3(n+2)}.
\label{U}
\end{equation}
The potential (\ref{U}) has one maximum at
$a=a_e=\sqrt{3/(2\Lambda)}$ such that
\begin{equation}
U_e\equiv
U(a_e)=\frac{s^23^{n+1}}{2^{n+1}\Lambda^{n+1}(n+1)(n+2)},
\label{Ue}
\end{equation}
so $U_e>0$ if (i) $n<-2$ or (ii) $n>-1$. The first case allows us
to solve the flatness and "Lambda" problems. The surplus dividend
of the model  is the presence of finite time region during which
universe has accelerated expansion.
\subsection{\label{sec:level2}The semiclassical tunneling probability
in VSL models with $n<-2$: the case $E\ll U_e$} One can choose
$n=-2-m$ with $m>0$. Such a substitution gives us the potential
(\ref{U}) in the form
\begin{equation}
U_m(a)=\frac{s^2}{a^{2(m+1)}}\left(\frac{2\Lambda
a^2}{3m}-\frac{1}{m+1}\right). \label{Um}
\end{equation}

The equation (\ref{equation}) is quite similar to equation for the
particle of energy $E$ that is moving in potential (\ref{Um}),
hence the universe in quantum cosmology can start at $a\sim 0$,
expand to the maximum radius $a'_i$ and then tunnel through the
potential barrier to the regime of unbounded expansion with
"initial" value $a=a_i$. The semiclassical tunneling probability
can be estimated as
\begin{equation}
\emph{P}\sim\exp\left(-2\int_{a'_i}^{a_i} {\mid {\tilde p(a)}\mid}
da\right), \label{P}
\end{equation}
with
$$
{\mid {\tilde p(a)}\mid}=\frac{c^2(t)}{8\pi G\hbar}\mid
p(a)\mid,\qquad \mid p(a)\mid=\sqrt{U_m(a)-E},
$$
where $E\le U_e$. It is  convenient to write $E=U_e\sin^2\theta$,
with $0<\theta<\pi/2$.

For the case $E\ll U_e$ one can choose
\begin{equation}
\displaystyle{ a'_i\sim a_1=\sqrt{\frac{3m}{2(m+1)\Lambda}},\qquad
a_i\sim
\sqrt{\frac{3}{2\Lambda}}\left(\frac{\sqrt{m+1}}{\sin\theta}\right)^{1/m}
,} \label{ai-ai}
\end{equation}
 and evaluate the integral (\ref{P}) as
\begin{equation}
\displaystyle{
\emph{P}\sim\exp\left(\frac{-s^3\Lambda^{2+3m/2}I_m(\theta)}{4\pi
G\hbar}\right),} \label{P1}
\end{equation}
where
\begin{equation}
I_m(\theta)=\int_{z'_i(\theta)}^{z_i(\theta)} dz
z^{-5-3m}\sqrt{\frac{2z^2}{3m}-\frac{1}{m+1}}, \label{Im}
\end{equation}
with
$$
 z'_i(\theta)=\sqrt{\frac{3m}{2(m+1)}},\qquad
z_i(\theta)=\sqrt{1.5}\left(\frac{(m+1)^{1/2}}{\sin\theta}\right)^{1/m}.
$$
The integral (\ref{Im}) can be calculated for the $m\in Z$. For
example
\begin{widetext}
$$
I_1(\theta)=\frac{\sqrt{3}}{17010}\left(3+\cos^2\theta\right)\left(191-78
\cos^2\theta+15 \cos^4\theta\right)\sqrt{6+2 \cos^2\theta}\sim
0.148+O(\theta^6),
$$
\end{widetext}
$$
I_2\sim 0.025,\qquad I_3\sim 0.007,\qquad I_4\sim 0.002,
$$
and so on. One can further show that $I_m(\theta)>0$ at
$0<\theta\ll 1$. Thus, it is easy to see from (\ref{P1}) that the
semiclassical tunneling probability $\emph{P}\to 0$ for large
values of $\Lambda>0$ and $\emph{P}\to 1$ at $\Lambda\to 0$.

Note, that the case $c$=const can be obtained by substitution
$m=-2$ into the (\ref{P1}). Not surprisingly, this case will get
us the well known result $\emph{P}\sim\exp(-1/\Lambda)$ (see
\cite{7}).
%%%%%%%%%%%%%%%%%%%%%%%%%%%%%%%%%%%%%%%%%%%%%%%%%%%%%
\subsection{\label{sec:level3}The semiclassical tunneling probability with $n<-2$ and
$n>-1$}

In the case of general position the semiclassical tunneling
probability with $n=-2-m$ has the form
\begin{equation}
\emph{P}_m\sim \exp\left(-\frac{s^3\Lambda^{(3m+4)/2}}{4\pi G\hbar
3^{(m+1)/2}\sqrt{m(m+1)}}\int_{z'_i}^{z_i}
\frac{dz\sqrt{F_m(z,\theta)}}{z^{3m+5}}\right), \label{Pm}
\end{equation}
where
\begin{equation}
F_m(z,\theta)=-2^{m+1}\sin^2\theta\, z^{2(m+1)}+2\times 3^m (m+1)
z^2-m 3^{m+1}, \label{Fm}
\end{equation}
$z$ is dimensionless quantity and  $z'_i$, $z_i$ are the turning
points, i.e. two real positive solutions of the equation
$F_m(z,\theta)=0$ for the given $\theta$ (it is easy to see that
the equation $F_m(z,\theta)=0$ does have two such solutions at
$0<\theta<\pi/2$).

If $m$ is the natural number then the expression (\ref{Pm}) has a
more simple form. For example
$$
\emph{P}_1\sim\exp\left(-\frac{s^3 \Lambda^{7/2}\sin\theta }{6\pi
G\hbar\sqrt{2}}\int_{z'_i}^{z_i}\frac{dz}{z^8}\sqrt{(z^2-{z'_i}^2)(z_i^2-z^2)}\right),
$$
with
$$
z'_i=\frac{\sqrt{3}}{2\cos(\theta/2)},\qquad
z'_i=\frac{\sqrt{3}}{2\sin(\theta/2)}.
$$
This expression can be calculated exactly:
$$
\emph{P}_1\sim\exp\left(-\frac{s^3\Lambda^{7/2}\sin\theta
J(\theta)}{6\sqrt{2}\pi G\hbar}\right),
$$
with
\begin{widetext}
$$
J(\theta)=\frac{1}{105}\left(\frac{2\sin(\theta/2)}{\sqrt{3}}\right)^5
\left[\frac{8\lambda^4-13\lambda^2+8}{\cos^2(\theta/2)}
\Pi\left(\mu^2;\frac{\pi}{2}{\backslash}\arcsin\mu\right)-
2\left(2\lambda^4-\lambda^2+2\right) \textrm{K}(\mu^2)\right],
$$
\end{widetext}
where $\mu^2=\cos\theta/\cos^2(\theta/2)$,
$\lambda=\cot(\theta/2)$, $\Pi$ and $\textrm{K}$ are complete
elliptic integral of the first and the third kinds correspondingly
\cite{8}.

Similarly, $\emph{P}\sim\exp(-S)$, with
$$
S=\frac{s^3 \Lambda^5\sin\theta }{18\pi
G\hbar}\int_{z'_i}^{z_i}\frac{dz}{z^{11}}\sqrt{(z^2+z_1^2)(z^2-{z'_i}^2)(z_i^2-z^2)},
$$
where
$$
\begin{array}{cc}
z_1=\sqrt{\frac{3}{\sin\theta}\cos\left(\frac{\theta}{3}-\frac{\pi}{6}\right)},\qquad
z'_i=\sqrt{\frac{3}{\sin\theta}\sin\frac{\theta}{3}},\\
\\
z_i=\sqrt{\frac{3}{\sin\theta}\cos\left(\frac{\theta}{3}+\frac{\pi}{6}\right)},
\end{array}
$$
and so on.

Therefore the probability to obtain (via quantum tunneling through
the potential barrier) the universe in the regime of unbounded
expansion is strongly suppressed for large values of $\Lambda$ and
small values of the initial scale factor $a_i
=\sqrt{3}/(2\sin(\theta/2)\sqrt{\Lambda})$. In other words,
overwhelming majority of universes which are nascent via quantum
tunneling through the potential barrier (\ref{U}) have large
initial scale factor and small value of $\Lambda$.

%%%%%%%%%%%%%%%%%%%%%%%%%%%%%%%%%%%%%%%%%%%%%%%

Now, let us consider the case (ii), when $n>-1$. The "quantum
potential" has the form
\begin{equation}
U(a)=s^2a^{2m}\left(\frac{1}{m}-\frac{2\Lambda
a^2}{3(m+1)}\right), \label{UU}
\end{equation}
where $m=n+1>0$. The points of intersection with the abscissa axis
$a$ are $a_0=0$ and $a_1=\sqrt{3(m+1)/2\Lambda m}$. Choosing $E=0$
in equation (\ref{equation}) and substituting (\ref{UU}) into the
(\ref{P}) we get
$$
\emph{P}\sim\exp\left(-\frac{s^3\Lambda^{(1-3m)/2}}{4\pi
G\hbar}\int_0^{z_1}
z^{2m-2}\sqrt{\frac{1}{m}-\frac{2z^2}{3(m+1)}}\,dz\right),
$$
with $z_1=\sqrt{3(m+1)/2m}$ (The starting value $z=0$ means that
the Universe tunneled from "nothing" to a closed universe of a
finite radius $a_1=z_1/\sqrt{\Lambda}$.). Thus, we have the same
effect as if $0<m<1/3$.

%%%%%%%%%%%%%%%%%%%%%%%%%%%%%%%%%%%%%%%%%%%%%%%%%%%%

\subsection{\label{sec:level4}Peculiar cases with $n=-1$ and $n=-2$}

At last,  lets consider the cases of $n=-1$ and $n=-2$. The
formula (\ref{Pm}) is not valid in these cases ($m=-1$ and $m=0$)
so we shall consider these models separately.

If $n=-1$ ($m=-1$) then
$$
\rho=\frac{M}{a^4}+\frac{\Lambda s^2}{8\pi Ga^2}-\frac{3s^2}{4\pi
G a^4}\log\frac{a}{a_*},
$$
therefore
\begin{equation}
U(a)=s^2\left(2\log\left(\frac{a}{a_*}\right)-\frac{2a^2\Lambda}{3}+1\right),
\label{U-1}
\end{equation}
where $a_*$ is constant and $[a_*]$=cm. The potential (\ref{U-1})
has one maximum at $a=a_e=\sqrt{3/(2\Lambda)}$ such that
$U_e=U(a_e)=2s^2\log(a_e/a_*)$, so if $a_e>a_*$ then $U_e>0$. We
choose $a_*=\Lambda^{-1/2}$. This gives us $U_e=0.41 s^2>0$. For
the case $E\ll U_e$ the semiclassical tunneling nucleation
probability is
\begin{equation}
\begin{array}{cc}
\displaystyle{
\emph{P}_{_{-1}}\sim\exp\left(-\frac{s^3\sqrt{\Lambda}}{4\pi
G\hbar}\int_{z'_i}^{z_i}\frac{dz}{z^2}\sqrt{\log
z^2-\frac{2z^2}{3}+1}\right)\sim}\\
\\
\displaystyle{\sim\exp\left(-\frac{s^3\sqrt{\Lambda}}{10\pi
G\hbar}\right),} \label{Pr1}
\end{array}
\end{equation}
where the turning points are $z'_i=0.721$, $z_i=1.812$. As we can
see from the (\ref{Pr1}), when $n=-1$ we receive the
aforementioned effect again.

If $n=-2$ ($m=0$) then
$$
\rho=\frac{M}{a^4}+\frac{s^2\Lambda}{2\pi
Ga^4}\log\left(\frac{a}{a_*}\right)+\frac{3s^2}{4\pi Ga^6}.
$$
We choose $a_*=1/(\alpha\sqrt{\Lambda})$, where $\alpha$ is a
dimensionless quantity. Thus
\begin{equation}
U(a)=-s^2\left(\frac{1}{a^2}+\frac{4\Lambda}{3}\log\left(\alpha
a\sqrt{\Lambda}\right)+\frac{\Lambda}{3}\right). \label{U-2}
\end{equation}
The maximum of potential (\ref{U-2}) is located at the same point
$a_e$ and
$$
U_e=-\frac{s^2\Lambda}{3}\left(3+\log\left(\frac{9\alpha^4}{4}\right)\right).
$$
Therefore, $U_e>0$ if $\alpha<2{\rm e}^{-3/4}/\sqrt{6}\sim 0.386$.
Choosing $\alpha=0.286$ and $E\ll U_e$ gets us the turning points
$z'_i\sim 0.77$ and $z_i\sim 2.391$.

At last, the semiclassical tunneling nucleation probability is
$$
\begin{array}{cc}
\displaystyle{\emph{P}_{0}\sim\exp\left(-\frac{s^3\Lambda^2}{4\pi
G\hbar}\int_{z'_i}^{z_i}\frac{dz}{z^4}\sqrt{-\frac{1}{z^2}-\frac{4}{3}\log(\alpha
z)-\frac{1}{3} }\right)\sim}\\
\\
\displaystyle{\exp\left(-\frac{0.084 s^3\Lambda^2}{\pi
G\hbar}\right)}.
\end{array}
$$
\section{An existence of ground states for singular potentials }

It is interesting to ask: what can be said about the value of
$U(a)$ when $a=0$?  What can be the possible meaning of the
potential which at $a\to 0$ is unbounded from below? It seems that
such universe is able to just roll down towards small values of
$a$ (where the potential is tending to minus infinity) instead of
any tunneling to large values.

This situation can in fact be alleviated if the considered
potential $U(a)$ has the ground state. Indeed, one can imagine the
fictitious particle with some energy and coordinate $a(t)$ in the
potentials (\ref{U}) rolling down towards small values of $a$. The
main problem is: whether the quantum mechanical energy spectrum of
$U(a)$ is unbounded below? If not, then it does admit the ground
state and hence can have the physical meaning.

To find such a potential lets suppose that our fictitious particle
is located in a small region $a$ near the singularity $a=0$, with
the momentum $P$. We will consider the case of arbitrary $w$. In
this case the flatness problem can be solved in early universe by
an interval of VSL evolution if $n < n_{_{fl}}(w)=-(1+3w)/2$,
whereas the problem of $\Lambda$-term can be solved only if
$n<n_{_\Lambda}(w)=n_{_{fl}}(w)-1=-3(w+1)/2$.

One can use the Heisenberg uncertainty relation as
\begin{equation}
P\,a\sim \left(8\pi G\hbar\right)^{(1-n_{_{fl}}(w))/2}
c^{(3n_{_{fl}}(w)-1)/2}. \label{Heisen}
\end{equation}
Using (\ref{Heisen}), and (\ref{frid}) (or (\ref{equation})) one
get for the $a\to 0$
$$
\begin{array}{cc}
\displaystyle{ E=P^2+U(a)\to}\\
\\
\displaystyle{\to \frac{Z^2}{a^{2-n(3n_{_{fl}}(w)-1)}}+\frac{s^2
n_{_{fl}}(w)}{(n_{_{fl}}(w)-n)a^{2(n_{_{fl}}(w)-n)}}},
\end{array}
$$
where $Z^2=\left(8\pi G\hbar\right)^{1-n_{_{fl}}}
s^{3n_{_{fl}}-1}$, and
\begin{equation} n<n_{_{fl}}(w)<0.
\label{nerv0}
\end{equation}
Therefore the energy spectrum will be bounded below if
\begin{equation}
\left(3n+2\right)\left(n_{_{fl}}(w)-1\right)<0. \label{nervy}
\end{equation}
and (\ref{nerv0}) are valid. This situation is represented
graphically on the Fig. 2.
%%%%%%%%%%%%%%%%%%%%%%%%%%%%%%%%%%%%%%%%%%%%%%%%%
\begin{figure}
\centering\leavevmode\epsfysize=5.5cm \epsfbox{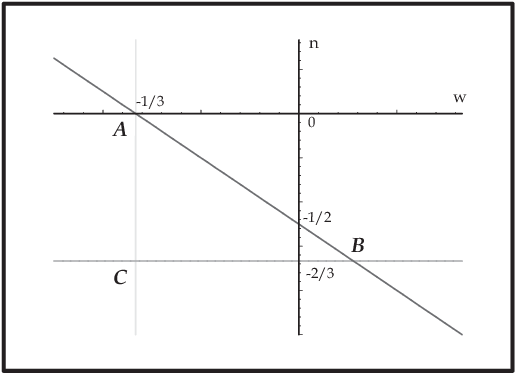}
\newline
\caption{\label{fig:epsart} The ground state exists for $w$ and
$n$ from the interior of the triangle ABC.}
\end{figure}
%%%%%%%%%%%%%%%%%%%%%%%%%%%%%%%%%%%%%%%%%%%%%%%%%
It is easy to see that the conditions of ground state existence
can be satisfied if $n_{_\Lambda}(w)<n<n_{_{fl}}(w)$. (The
additional restriction $w>-1/3$ is just a condition of existence
of maximum of the potential $U(a)$. For more detailed examination
of the general case see \cite{VSL}). In the case of universe filed
with radiation (as above) one get $-2<n<-1$.
\section{Instantons}
If we are going to describe the quantum nucleation of universe we
should find the instanton solutions, simply putted as a stationary
points of the Euclidean action. The instantons give a dominant
contribution to the Euclidean path integral, and that is the
reason of our interest in them.

First at all, lets consider the $O(4)$-invariant Euclidean
spacetime with the metric
\begin{equation}
ds^2=c^2(\tau)d\tau^2+a^2(\tau)\left(d\psi^2+\sin^2\psi
d\Omega_2^2\right). \label{metric}
\end{equation}
In the case $c={\rm const}$ one can construct the simple
instantons, which are the $O(5)$ invariant four-spheres. Then one
can introduce the scalar field $\phi$, whose (constant) value
$\phi=\phi_0$ is chosen as the one providing the extremum of
potential $V(\phi)$. The scale factor will be
$a(\tau)=H^{-1}\sin\,H\tau$ and after the analytic continuation
into the Lorentzian region one will get the de Sitter space or
inflation. Many other examples of non-singular and singular
instantons were presented in \cite{Linde-98}

Now, lets consider the VSL model with scalar field. Here we get
the following Euclidean equations:
\begin{equation}
\begin{array}{l}
\displaystyle{
\phi''+3\frac{a'}{a}\phi'=\frac{c^2V'}{\phi'}+\frac{c^5c'(\Lambda
a^2-3)}{4\pi G
a^2\phi'}+\frac{2\phi'c'}{c}-\frac{2cVc'}{\phi'},}\\
\\
\displaystyle{ \left(\frac{a'}{a}\right)^2=\frac{8\pi
G}{3c^4}\left(\frac{\phi'^2}{2}-c^2V\right)+\frac{c^2}{a^2}-\frac{\Lambda
c^2}{3}},
 \label{ins-1}
\end{array}
\end{equation}
where primes denote derivatives with respect to $\tau$.

At the next step we represent the potential $V$ in factorized form
\begin{equation}
V=F(a)U(\phi). \label{ins-2}
\end{equation}
Indeed, lets for example consider the power-low potential
$\sim\phi^k$. If the coupling $\lambda$ is dimensionless one then
we get
$$
V\sim \frac{\lambda}{\hbar} G^{k/2-2} c^{7-2k}\phi^k.
$$
Since $c=sa^n$ then in the simplest case we come to (\ref{ins-2}).

Let $\phi=\phi_0={\rm const}$ be solution of the (\ref{ins-1}).
(Note, that we don't require the $\phi_0$ to be the extremum of
potential. Another interesting point is the possibility to consider nontrivial soliton solution of classical field equations, see for example \cite{BLP}.) Using the first equation of system (\ref{ins-1}) and
(\ref{ins-2}) we get the equation for the $F(a)$,
\begin{equation}
\frac{dF(a)}{da}-\frac{2n}{a}F(a)=\frac{3ns^4}{4\pi GU_0
}a^{4n-3}-\frac{ns^4\Lambda}{4\pi G U_0}a^{4n-1} , \label{ins-3}
\end{equation}
where $U_0=U(\phi_0)={\rm const}$. The integration of the
(\ref{ins-3}) results in
\begin{equation}
F(a)=a^{2n}\left(C-\frac{3ns^4}{8\pi
G(1-n)U_0}a^{2(n-1)}-\frac{s^4\Lambda}{8\pi G U_0} a^{2n}\right),
\label{ins-4}
\end{equation}
where $C$ is the constant of integration and by assumption $n\neq -1$ and $n\neq 0$. Substitution of
(\ref{ins-4}) into the second equation of the system (\ref{ins-1})
transforms it into the the model of nonlinear oscillator the
integration of which result in
\begin{equation}
 \frac{a'^2}{2}+u(a)=0,
\label{ins-5}
\end{equation}
where
\begin{equation}
u(a)=\frac{\omega^2 a^2}{2}-\frac{s^2 a^{2n}}{2(1-n)},
\label{ins-6}
\end{equation}
with $\omega^2=8\pi G U_0 C/(3s^2)$ and with the choice $C> 0$
made. We can see that for $c={\rm const}$ (i.e. $n=0$) the
equation (\ref{ins-6}) turns out to be just the usual harmonic
oscillator and we come to the well-known $O(5)$ solution (but in
this case $\phi_0$ must be the stationary point of $V$).
%%%%%%%%%%%%%%%%%%%%%%%%%%%%%%%%%%%%%%%%%%%%%

The equation (\ref{ins-5}) naturally describes the ''movement of a classical
particle'' with zero-point energy in mechanical potential
(\ref{ins-6}). Depending on value of $n$ this potential can take one of four distinct forms
(excluding the well-known classical case $n=0$, which lies beyond the scoop of this article).

\textbf{Case 1: $n<0$.} Potential $u(a)$ has the form, depicted on the Fig. 3.
%%%%%%%%%%%%%%%%%%%%%%%%%%%%%%%%
\begin{figure}
\centering\leavevmode\epsfysize=5.5cm \epsfbox{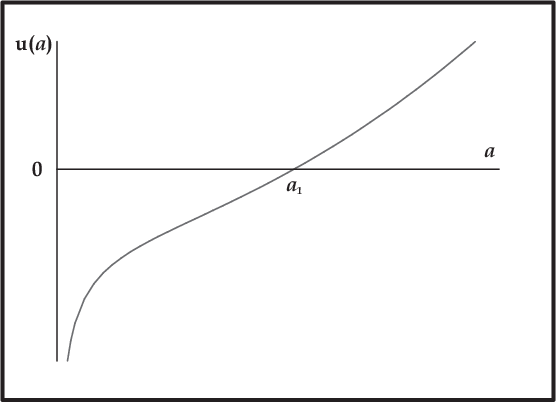}
\newline
\caption{\label{fig:epsart} The potential $u(a)$. Instanton exists
in region $0\le a\le a_1$ but only if $n>-1/5$.}
\end{figure}
%%%%%%%%%%%%%%%%%%%%%%%%%%%%%%%%
Here we have one Euclidean ($0\le a\le a_1$) and one Lorentzian
($a>a_1$) regions where
\begin{equation}
a_1=\left(\frac{s}{\omega\sqrt{1-n}}\right)^{1/(1-n)}.
\label{Intersection}
\end{equation}
On the bound between Euclidean and Lorentzian regions ($a=a_1$)
we have $a'=0$.

This mechanical potential is unbounded from below at $a\to
0$. With this in mind, we'll have to ascertain that the Euclidean
action for our solution will stay finite. The gravitation action
has the form
$$
S_{{\rm grav}}=-\int d^4x \frac{c^3}{8\pi G}\sqrt{g} R.
$$
We are using the dimensionless variables $x^0=c_0\tau/a_0$,
$x^1=\psi$ and so on. Calculating $R$ we get
\begin{equation}
R=\frac{6}{c_0^2a^2}\left[c_0^2-\left(\frac{a_0}{a}\right)^{2n}\left((1-n)a'^2+aa''\right)\right],
\label{Ricci}
\end{equation}
so we do have the potential divergence at $a=0$. Multiplying
(\ref{Ricci}) on the $\sqrt{g}$ and $c^3$ and using the equation
of motion we get the expression:
\begin{equation}
R\sqrt{g} c^3\sim
6c_0\left((2-n)\omega^2\frac{a^{2n+3}}{a_0^{2n-1}}-\frac{nc_0^2}{1-n}\frac{a^{4n+1}}{a_0^{4n-1}}\right),
\label{Ricci-1}
\end{equation}
where the most dangerous multiplier factor is $a^{1+4n}$. But if
$-1/4 \le n < 0$ then the Euclidean action becomes finite and
therefore, we end up with the legitimate gravitation instanton. In
a similar manner, using (\ref{ins-2}) and (\ref{ins-4}) we get for
the scalar field (in dimensionless $x^{\mu}$):
%\begin{widetext}
$$
\begin{array}{l}
\displaystyle{\sqrt{g}V_0\sim\frac{c_0a_0^{1-3n}}{8\pi
G}(3\omega^2
a^{3(1+n)}+}\\
\\
\displaystyle{+\frac{3nc_0^2a^{1+5n}}{(n-1)a_0^{2n}}-\frac{\Lambda
c_0^2a^{3+5n}}{a_0^{2n}})}
\end{array}
$$
%\end{widetext}
therefore the instanton exists for $n>-1/5$. This demand is more
powerful than what we got for the gravitation instanton where
$n>-1/4$ (see (\ref{Ricci-1})).

\textbf{Case 2. $0<n<1$.} Here the potential $u(a)$ suffers no singularity at
$a=0$, but $u(0)=0$. Also this potential has a minimum at
$$
a_0=\left(\frac{s}{\omega}\sqrt{\frac{n}{1-n}}\right)^{1/(1-n)},
$$
and is equal to zero at (\ref{Intersection}), hence, once again creating one
Euclidean and one Lorentzian regions, separated by (\ref{Intersection}).

\textbf{Case 3. $n=1$.} This case is somehow special, since for such $n$ the solution
of (\ref{ins-3}) shall be
$$
F(a)=a^{2n}\left(C-\frac{3s^4}{4\pi
G U_0} \ln{a}-\frac{s^4\Lambda}{8\pi G U_0} a^{2}\right),
$$
instead of (\ref{ins-4}), and hence, the equation of (\ref{ins-6}) shall be substituted by
\begin{equation}
u(a)=a^2 \left(\frac{\omega^2}{2}-s^2 \ln{a} \right).
\label{inss-1}
\end{equation}
It is easy to see that this function has two zeros (at $a_1=0$ and $a_2=\exp(\frac{\omega^2}{2 S^2})$),
is strictly positive at the interval ($a_1$, $a_2$) and strictly negative outside of it. Therefore, this case doesn't
admit the instanton.

\textbf{Case 4. $n>1$.} The potential $u(a)$ is strictly positive. The instanton doesn't exist either.

Both of a newly founded solutions possess only $O(4)$ invariance just like
Hawking-Turok instanton (so the Euclidean region is a deformed
four sphere) but, unlike to it, they are all non-singular. Note that if the
value $a$ is sufficiently large then one can neglect the second
term in (\ref{ins-6}) (after the analytic continuation into the
Lorentzian region) therefore, as in the case of  the usual $O(5)$
instanton, one can  get  the de Sitter universe, i.e. the
inflation.
%%%%%%%%%%%%%%%%%%%%%%%%%%%%%%%%%%%%%

The equation (\ref{ins-5}) has no terms with $\Lambda$. In
other words, the scale factor $a(\tau)$ doesn't depend on the
value $\Lambda$ (although being dependant on the $U_0$).
Therefore, the full Euclidean action $S_{_E}=S_{{\rm
grav}}+S_{{\rm field}}$ has the form,
$$
S_{_E}=S_0-\Lambda S_1,
$$
where $S_0$ and $S_1$ are both independent of the $\Lambda$.
Returning to what has been said in Introduction, there exist three
common ways to describe the quantum cosmology: the
Hartle-Hawking wave function $\exp(-S_{_E}/\hbar)$, the Linde wave
function $\exp(+S_{_E}/\hbar)$ and the tunneling wave function. In
the second Section we have been working with the tunneling wave
function. In case of instantons situation becomes slightly different.
If $S_1>0$ then (as a first, tree semiclassical approximation) we
should choose the Linde wave function, whereas for the case
$S_1<0$ the Hartle-Hawking wave function seems more naturally.

In conclusion, we note that another choice of $C$ ($C<0$ and
$C=0$) eliminates any possible instantons.

\section{Discussion}

VSL models contain both some of the promising positive features
\cite{1}  and some shortcomings and unusual (unphysical?) features
as well \cite{3}. But, as we have shown, application of the VSL
principle to the quantum cosmology indeed results in amazing
previously unexpected observations. The first observation is that
the semiclassical tunneling nucleation probability in VSL quantum
cosmology is quit different from the one in quantum cosmology with
$c$=const. In the first case this probability  can be strongly
suppressed for large values of $\Lambda$ whereas in the second
case it is strongly suppressed for small values of $\Lambda$. This
is interesting, although we still can't say that VSL quantum
cosmology definitely results in solution of the $\Lambda$-mystery.
The problem here is the validity the WKB wave function. And what
is more, throughout the calculations we have been omitting all
preexponential factors (or one loop quantum correction) which can
be essential ones near the turning points. Another troublesome
question is the effective potentials in VSL models, being
unbounded from below at $a\to 0$. The naive way to solve this
problem is to use the Heisenberg uncertainty relation to find
those potentials with the ground state. However, this is just a
crude estimation. To describe the quantum nucleation of universe
we have to find the instanton solution which, being a stationary
point of the Euclidean action, gives the dominant contribution to
the Euclidean path integral. As we have seen, such solutions indeed
exist in VSL models. Those instantons are $O(4)$ invariant, are
non-singular, and provide an inflation as well. They describe the quantum
nucleation of universe from ''nothing'' and, what is more, upon usage of these
solutions we can obtain the probability of a nucleation which is
suppressed for large value of $\Lambda$ (as in see Sec. II) using
either Linde or Hartle-Hawking wave function.

Note, that we can weaken the condition $n>-1/5$ to obtain a
singular instanton suffering the integrable singularity (i.e. such
that the instanton action will be finite) in the way of the
Hawking-Turok instanton. However, there exist some arguments
\cite{Vilenkin-98}, that such singularities, even being
integrable, still lead to serious problems with solutions.

In conclusion, we note that obtained instantons both have a free parameter
($\omega^2$) so we are free to use the anthropic approach to find
the most probable values of $\Lambda$ too.

\begin{acknowledgments}
After finishing the first version of this work, we learned that
T.Harko, H.Q.Lu, M.K.Mak and K.S.Cheng \cite{Harko}, have
independently considered the VSL tunneling probability in both
Vilenkin and Hartle-Hawking approaches.  The interesting
conclusion of their work is that at zero scale factor the
classical singularity is no longer isolated from the Universe by
the quantum potential but instead classical evolution can start
from arbitrarily small size. In contrast to \cite{Harko}, we
attract attention to the problem of $\Lambda$-term and instantons
in VSL quantum cosmology.

We'd like to thank Professor S. Odintsov for his encouraging
friendly support, Professor T. Harko for useful information about
the article \cite{Harko}, Professor Pedro F. Gonz\'alez-D\'iaz,
Prado Mart\'in-Moruno and Salvador Robles-Per\'ez for usefull
discussion. One of us (A.Yu.) thanks the Instituto de
Matem\'aticas y F\'izica Fundamental (Madrid, Spain) for
hospitality.

\end{acknowledgments}

\appendix

\section{The case $n>0$}

Usually those of authors, who apply an idea of variability of the light speed to cosmology, are restricting themselves to studying the models with negative values of $n$ (upon doing so they became able to solve some of the cosmological problems without involvement of inflation). On the other hand, as we have shown in the last section, the models with positive $n$ involve a nonsingular instanton, and hence are sensible from the point of view of quantum cosmology. Moreover, as we shall see, such models can also describe \textbf{our} observed universe; in particular, they provides
a solution to the flatness problem and describe the present acceleration of universe. Finally, the models with positive $n$ can naturally result in Big Rip, as will be shown below.

\subsection{\label{sec:level2}The dust case:  flatness and acceleration }
The integration of (\ref{frid}) (with assumption $p=wc^2\rho$) results in
\begin{equation}
\rho=\frac{M}{a^{3(w+1)}}+\frac{3nks^2a^{2(n-1)}}{4\pi
G(3w+2n+1)}-\frac{ns^2\Lambda a^{2n}}{4\pi G(3w+2n+3)},
\label{rho_new}
\end{equation}
where $M$=const.  The pressure is
$$
p=\frac{ws^2M}{a^{3w+3-2n}}+\frac{3kns^4wa^{4n-2}}{4\pi
G(3w+2n+1)}-\frac{s^4n\Lambda wa^{4n}}{4\pi G(3w+2n+3)}.
$$
Therefore
\begin{equation}
\begin{array}{l}
\displaystyle{ \frac{\ddot a}{a}=-\frac{4\pi
GM(3w+1)}{3a^{3(w+1)}}-\frac{s^2nk(3w+1)a^{2(n-1)}}{3w+2n+1}}+\\
\displaystyle{ +\frac{s^2\Lambda(w+1)(n+1)a^{2n}}{3w+2n+3}.}
\label{ddota}
\end{array}
\end{equation}
An equation on the Hubble constant $H={\dot a}/{a}$ is
\begin{equation}
H^2=\frac{8\pi
GM}{3a^{3(w+1)}}-\frac{(3w+1)ks^2a^{2(n-1)}}{3w+2n+1}+\frac{\Lambda
s^2(w+1)a^{2n}}{3w+2n+3}. \label{H2}
\end{equation}

If $w=0$ then equations (\ref{ddota}) and (\ref{H2}) are reduced
to
\begin{equation}
\begin{array}{l}
\displaystyle{
\frac{\ddot a}{a}=-\frac{4\pi GM}{3a^3}-\frac{s^2nka^{2(n-1)}}{2n+1}+\frac{s^2\Lambda (n+1)a^{2n}}{2n+3},}\\
\\
\displaystyle{ H^2=\frac{8\pi
GM}{3a^3}-\frac{ks^2a^{2(n-1)}}{2n+1}+\frac{\Lambda
s^2a^{2n}}{2n+3}.}
\end{array}
\label{sys}
\end{equation}
The modern observations result in
\begin{equation}
\frac{{\ddot a}_0}{a_0}\sim\frac{7H_0^2}{10}, \label{obs}
\end{equation}
where $H_0=h\times 0.324\times 10^{-19}$ ${\rm s}^{-1}$, with
$h=45 \div 75$, $a_0\sim 10^{28}$ cm. Thus we have
\begin{equation}
\frac{{\ddot a}_0}{a_0}=-\frac{4\pi
GM}{3a_0^3}-\frac{nkc_0^2}{a_0^2(1+2n)}+\frac{c_0^2\Lambda(n+1)}{3+2n},
\label{uskor}
\end{equation}
and
\begin{equation}
H_0^2=-\frac{kc_0^2}{(1+2n)a_0^2}+\frac{\Lambda
c_0^2}{3+2n}+\frac{8\pi G M}{3a_0^3}. \label{H0}
\end{equation}
Using (\ref{obs}), (\ref{uskor}) and (\ref{H0}) we
get to
\begin{equation}
M=\frac{3a_0\left[a_0^2H_0^2(1+2n)(3+10n)+10kc_0^2\right]}{40\pi
G(1+2n)(3+2n)}, \label{M}
\end{equation}
\begin{equation}
\Lambda=\frac{5kc_0^2+12H_0^2a_0^2}{5a_0^2c_0^2}, \label{L}
\end{equation}
hence, concluding that such model are indeed able to describe our
observable universe.

Let $n$ be positive albeit very small quantity: $n\ll 1$
(and $w=0$). In this case the equations (\ref{rho_new}),
(\ref{M}),(\ref{sys}) will be reduced to
$$
\rho=\frac{M}{a^3}-\frac{ns^2\Lambda a^{2n}}{12\pi
G}+\frac{3nks^2}{4\pi Ga^2},
$$
$$
M=\frac{a_0(3H_0^2a_0^2+10kc_0^2)}{40\pi G},
$$
\begin{equation}
H^2=\frac{8\pi GM}{3a^3}-\frac{ks^2}{a^2}+\frac{\Lambda
s^2a^{2n}}{3}, \label{HH}
\end{equation}
\begin{equation}
\frac{\ddot a}{a}=-\frac{4\pi
GM}{3a^3}-\frac{s^2nk}{a^2}+\frac{\Lambda s^2a^{2n}}{3}.
\label{vot}
\end{equation}
It easily follows from the (\ref{vot}) that the evolution of the universe can be divided into three stages:
\newline
\newline
Stage I. For values of $a$ that are close enough to the initial singularity (i.e. $a\to 0$) the first term in
(\ref{vot}) dominates above the second and the third ones. The very young universe
has been a flat one. ~\footnote{Of course,
in close vicininear initial singularity it will be more correctly to consider
$w=1/3$ rather then $w=0$. But the result will be the same: the
universe was flat.}
\newline
\newline
Stage II. The second term in (\ref{vot}) dominates above the first
and third ones. Here we can't consider the universe as flat. But
we can do it on the last stage.
\newline
\newline
Stage III. The last term in (\ref{vot}) dominates above the first
and second ones. This stage is the one best fit to describe the universe we
resides in. In order to estimate the starting time of this phase we shall
use (\ref{vot}). In fact, it is easy to show that last stage will start only if
\begin{equation}
z<\frac{1}{\sqrt{n}}-1, \label{z}
\end{equation}
where $z=a_0/a-1$ is a red shift. Therefore, the modern astronomical observations can give us the upper bound on
$n$.

\subsection{\label{sec:level3}  Big Rip and Big Trip}
The integrals, appearing in (\ref{HH}) (and (\ref{vot})) in case of general position can be extremely hard to solve.  However, we can make everything a lot more easier, noting that at final stage both of the first terms in equation (\ref{HH}) can actually be omitted. Keeping this in mind, we come to
\begin{equation}
a(t)=\left(\frac{1}{ns}\sqrt{\frac{3}{\Lambda}}\right)^{1/n}\frac{1}{(t_*-t)^{1/n}},
\label{BR}
\end{equation}
here $t_*$=const. The value of $t_*$ (i.e. a time of a big rip
occurrence, see (\ref{ola-la})) can be obtained via integration of
unabridged equation (\ref{HH}). Thus, quite surprisingly, we get
the big rip (See \cite{Caldwell} about big rip in ''usual
cosmology'' with $c=$const).
%%%%%%%%%%%%%%%%%%%%%%%%%%%%%%%%%%%%%%%

{\bf Remark 3}. In the case of ''usual cosmology'' the big rip
singularity can exist because of phantom energy. There are few
ways to escape of future big rip singularity: (i) to consider
phantom energy just as some effective models (see \cite{444},
\cite{5555}, \cite{Aref'eva}); (ii) to use quantum effects to
delay the singularity \cite{Nojiri}; (iii) to use new time
variable such that the big rip singularity will be point at
infinity ($t\to\infty$) \cite{Multiverse}; (iv) to avoid big rip
via another cosmological ''Big'': big trip (see below).

In \cite{9} Pedro F. Gonz\'alez-D\'iaz had shown that phantom
energy can lead to an achronal cosmic future where the wormholes
become infinite before the arisement of the big rip singularity.
Soon after that, with the continuing accretion of a phantom
energy, any wormhole becomes the Einstein-Rosen bridge (see also
\cite{Faraoni}, \cite{Est}). Pedro F. Gonz\'alez-D\'iaz has
suggested that such objects can be used by an advanced
civilization as the means of escape from the big rip, but, via
usage of Bekenstein Bound we have shown it to be impossible due to
the very strong restrictions laid on the total amount of
information which can be sent through this bridge \cite{Privet}.
It is very interesting that in  VSL models with big rip the escape
\textbf{is} still possible.

To show this, we shall use Babichev, Dokuchaev and Eroshenko (BDE)
equation for the throat radius of a Morris-Thorne wormhole
\cite{111} in the dust universe with VSL during the last stage of
evolution with
$$
\rho\sim -\frac{s^2n\Lambda a^{2n}}{12\pi G}.
$$
In this case the BDE equation takes the form
\begin{equation}
\frac{1}{b(t)}=\frac{1}{b_0}+\frac{\pi D\sqrt{3\Lambda}}{6}\log
\left|\frac{t_*-t}{t_*-t_0}\right|, \label{b}
\end{equation}
where $b$ is a throat radius and $D$ is a positive constant. Hence, in assumption that
the equation of BDE remains correct in VSL models, we conclude that there exist a big
trip at $t=T<t_*$, where
$$
\displaystyle{ T=t_*-(t_*-t_0) \exp{\left(\frac{2}{\pi Db_0}\sqrt{\frac{3}{\Lambda}}\right)}},
$$
and $b_0$ is initial value of $b$. Now we can calculate the
horizon distance in such universe:
$$
R_c(t)=a(t)\int_t^{t_*} \frac{c(t')dt'}{a(t')}.
$$
Substituting (\ref{BR}) into this expression we get
$$
R_c(t)=\sqrt{\frac{3}{\Lambda}},
$$
at the last stage of evolution. Thus, we have no information bound
like the one taking place in models with phantom fields (where $R_c\to 0$
when $t\to t_*$) and so, the advanced civilization can in principle use such objects as a means of
escape from the big rip singularity.

Note that it will be interesting to use the method from the \cite{Idefenite} (dressing procedure) to construct exact solutions of the
Friedmann and Raychaudhuri system. Another interesting situation is connected with the quantum gravity
effects. Such effects take place at a scales $a\sim L_{Pl}$. Since
$$
L_{Pl}^2=\frac{G\hbar}{c^3}=\frac{G\hbar}{c_0^3}\left(\frac{a_0}{a}\right)^{3n},
$$
we can see that $L_{Pl}\to \infty$ at $a\to 0$ (or, in other words, the value of red shift
$z\to\infty$). It now follows, that the universe in vicinity of a Big Bang has been quantum. In
other words, in order to successfully describe the universe for large $z$ one needs the quantum
gravity theory. On the other hand, close to the big rip we get $L_{Pl}\to 0$!
Therefore, in VSL models, the big rip is a purely classical effect and we don't
need a quantum gravity (either string theory or loop quantum gravity) in order to
describe big rip in such universe.

%\section{Some notes about George F.R. Ellis and Jean-Philippe Uzan
%approach}

%
$$
{}
$$
%\newpage %Just because of unusual number of tables stacked at end
\bibliography{apssamp}% Produces the bibliography via BibTeX.

\vfill \eject

\end{document}